\documentclass[8.5pt,twoside,twocolumn]{article}
\usepackage[super,sort&compress,comma]{natbib} 
\usepackage[version=3]{mhchem}
\usepackage{balance}
\usepackage{times,mathptm}
\usepackage{amsmath}
\usepackage{secsty}
\usepackage{graphicx} 
\usepackage{lastpage}
\usepackage[format=plain,justification=raggedright,singlelinecheck=false,font=small,labelfont=bf,labelsep=space]{caption} 
\usepackage{fancyhdr}
\begin{document}

\thispagestyle{plain}
\fancypagestyle{plain}{
\renewcommand{\headrulewidth}{1pt}}
\renewcommand{\thefootnote}{\fnsymbol{footnote}}
\renewcommand\footnoterule{\vspace*{1pt}% 
\hrule width 3.4in height 0.4pt \vspace*{5pt}} 
\setcounter{secnumdepth}{5}

\makeatletter 
\renewcommand\@biblabel[1]{#1}            
\renewcommand\@makefntext[1]% 
{\noindent\makebox[0pt][r]{\@thefnmark\,}#1}
\makeatother 
\renewcommand{\figurename}{\small{Fig.}~}
\newcommand{\vshift}{f_{\text{shift}}}
\newcommand{\rcut}{r_{\text{cut}}}
\sectionfont{\large}
\subsectionfont{\normalsize} 

\renewcommand{\headrulewidth}{1pt} 
\renewcommand{\footrulewidth}{1pt}
\setlength{\arrayrulewidth}{1pt}
\setlength{\columnsep}{6.5mm}
\setlength\bibsep{1pt}

\twocolumn[
  \begin{@twocolumnfalse}
\noindent\LARGE{\textbf{Inverse design of simple pairwise interactions with low-coordinated 3D lattice ground states$^\dag$}}
\vspace{0.6cm}

\noindent\large{\textbf{Avni Jain\textit{$^{a}$}, Jeffrey R. Errington\textit{$^{b}$} and Thomas M. Truskett$^{\ast}$\textit{$^{a}$}}}\vspace{0.5cm}
%Please note that \ast indicates the corresponding author(s) but no footnote text is required. 

\noindent{\small{Article can be cited as \textbf{Soft Matter, 2013, DOI: 10.1039/C3SM27785B}}}
 \end{@twocolumnfalse} \vspace{0.6cm}
 ]

\noindent\textbf{We demonstrate that  inverse statistical mechanical optimization can be used to discover simple (e.g., short-range, isotropic, and convex-repulsive) pairwise interparticle potentials with three-dimensional diamond or simple cubic lattice ground states over a wide range of densities.}
\section*{}
\vspace{-1cm}

%Footnotes
%Please use \dag to cite the ESI in the main text of the article.
%If you article does not have ESI please remove the the \dag symbol from the title and the above footnotetext.

\footnotetext{\textit{$^{a}$~Department of Chemical Engineering, The University of Texas at Austin, Austin, TX 78712. E-mail: ttruskett@che.utexas.edu}}
\footnotetext{\textit{$^{b}$~Department of Chemical and Biological Engineering, University of Buffalo, The State University of New York, Buffalo, New York 14260-4200.}}

The properties of condensed phases are often linked to their structure. For example, heterogeneous materials with three-dimensional (3D) dielectric diamond morphologies can exhibit a photonic band gap~\cite{DIAphoton}, making them useful architectures for applications that range from lasers and sensors to solar cells. Although alternative methods for fabricating such materials have been recently introduced, considerable interest remains in understanding how to create systems that spontaneously self-assemble into structures with desirable properties. Moreover, since various aspects of the effective interactions between nanometer- to micron-scale particles can be tuned experimentally via modification of solution or particle properties~\cite{ Blaaderen_nature,Likos_review_2001}, the following fundamental materials design question becomes especially relevant. Which types of interparticle potentials provide a thermodynamic driving force for the particles to self-assemble into a given target lattice? 

Results from statistical mechanical theories, computer simulations, and experiments have produced valuable insights into how to design interparticle interactions for self-assembly into periodic structures. For example, it is widely appreciated that spherical particles with steeply repulsive interactions spontaneously assemble into highly-coordinated 3D structures~\cite{Pusey1986,PRLPusey1989}, such as the face-centered cubic (FCC) lattice, at sufficiently high particle concentrations. Interactions that  favor a targeted low-coordinated lattice ground state over other competing structures can also be designed by introducing specific types of complexity into the interparticle potential (e.g., multiple wells\cite{PREDiaWur,PRESC}, non-spherical particle shapes\cite{Glotzernature,Glotzerscience,oleggangACSNano}, or orientation-dependent ``patches'' on a particle surface~\cite{Zhangglotzer,designrulepatchy_glotzer,OgangNL2010,sciortinonatcom,patchyGKahl,PRLTkachenko,EGNoyaJCP,FRomanoJCP}), but those phases are generally stable over narrow ranges of thermodynamic conditions~\cite{EGNoyaJCP,FRomanoJCP}. On the other hand, whether interactions with considerably simpler functional forms can also produce targeted low-coordinated 3D ground states--stable over a wide range of densities--remains an interesting open question. 

Inverse statistical mechanical methods such as those pioneered in recent years by Torquato, Stillinger, and others~\cite{TorquatoRev,PREDiaWur,PRLFirst2DMR,2DMonotonicSM,JCP2dmonotonic,CohnPNAS,EdlundPRLmain} can be used to address this question. In fact, focusing on the specific case of two-dimensional systems, Marcotte, Stillinger, and Torquato~\cite{2DMonotonicSM,JCP2dmonotonic} have employed inverse design principles to discover isotropic, convex-repulsive potentials with low-coordinated square and honeycomb lattice ground states. In the present study, we build upon that insightful body of work to search for simple pair potentials with specific low-coordinated 3D lattice ground states that are stable over a wide range of density. We choose the symmetric Bravais simple cubic lattice and the asymmetric non-Bravais diamond lattice as our target structures. 

The pair potentials we consider in our optimization are isotropic, convex-repulsive, twice continuously differentiable, and short-ranged. They are described (in terms of a characteristic energy scale $\epsilon$ and length scale $\sigma$) by the functional form, \(V(r/\sigma)= \epsilon \{f(r/\sigma)+\vshift (r/\sigma)\} H[(\rcut-r)/\sigma\)]. Here, \(f(r/\sigma)\)--motivated by a recently introduced model\cite{fominmain}--is given by
\begin{equation} \label{eq:fomin}
  f\left(\frac{r}{\sigma}\right)=A\left(\frac{\sigma}{r}\right)^{n}+\sum_{j=1}^2\lambda_{i}\left\{1-\tanh\left[k_{j}\left(\frac{r_{}}{\sigma}-\delta_j\right)\right]\right\}
\end{equation}
 \(H\) is the Heaviside step function, and \(\vshift (r/\sigma)=X(r/\sigma)^2+Y(r/\sigma)+Z\). The constants $X$, $Y$, and $Z$ are implicit functions of the other parameters in the potential via the constraints, \(V(\rcut/\sigma)=V^{\prime}(\rcut/\sigma)=V^{\prime \prime}(\rcut/\sigma)=0\). In this study, we set $\rcut/\sigma=2.25$. As a result, $V(r/\sigma)/\epsilon$ depends on eight dimensionless parameters ($A$, $n$, $\lambda_{1}$, $k_{1}$, $\delta_{1}$, $\lambda_{2}$, $k_{2}$, $\delta_{2}$), but one of these is not free because we further require \(f(1)+\vshift(1)=V(1)/\epsilon=1\).    

We obtain optimized potential parameters for specific target structures using a standard simulated annealing algorithm (e.g., as described in Corana \emph{et al.}~\cite{Corana}). Our optimization goal is {\em to maximize the range of density over which the target lattice is the ground state} for the potential. One practical way of accomplishing this is to maximize the number $n$ of uniformly spaced densities within a wide range [\(\rho_{1},\rho_{1}+(m-1)\Delta\rho] \), where $0\le n \le m$, for which the zero-temperature chemical potential (molar enthlapy) of the target structure is lower than those of the competing periodic structures at the corresponding pressures. 

To get a more concrete sense of the optimization problem, consider the $i^{\text{th}}$ cycle at a given simulated annealing temperature $T_{\text{SA}}$. We first choose the eight primary potential parameters randomly, and subsequently solve for constants $X$, $Y$, and $Z$ via the aforementioned constraints. If these 11 parameters are inconsistent with a convex repulsive potential, we reject them and try again; otherwise, we rescale the potential (and hence the parameters $A$, $\lambda_{1}$, $\lambda_{2}$, $X$, $Y$, and $Z)$  by the constant factor required to ensure \(f(\sigma)+\vshift(\sigma)=1\). For the resulting $i^{\text{th}}$ trial potential, the zero-temperature pressure and chemical potential of the target lattice at density $\rho_1$ are computed. This chemical potential is then compared to that of all other lattices in the competitive pool of structures (discussed below) at the same pressure. Similar comparisons are also carried out at pressures corresponding to the other $m-1$ target lattice densities in the range of interest [$\rho_{1}+\Delta\rho$, $...$, $\rho_{1}+(m-1)\Delta\rho$]. The number of state points in this set for which the target structure has the minimum chemical potential in the competitive pool is  labeled $n_i$. The minimum chemical potential difference between the target and its competing structures considering all $m$ pressures is labeled \(\Delta\mu_i\) (a quantity which is negative if the target lattice is favored for at least one state point; i.e., if $n_i >0$). We define the simulated annealing energy for the $i^{\text{th}}$ trial potential E$^{i}_{\text{SA}}$ as 

\begin{equation}
{E^{i}_{\text{SA}}} = - n_i + {\text{H}} (\Delta\mu_i) \left[ n_i + \exp(\Delta\mu_i /\epsilon ) \right]
\end{equation}

In other words, the trial potential will be accepted as the $i^{\text{th}}$ cycle's pair potential with the standard Metropolis probability \({\text{min}}\left[1,\exp\left(\left({E^{i-1}_{\text{SA}}}\;-E^{i}_{\text{SA}}\right)/k_{\text{B}} T_{\text{SA}}\right)\right] \)  where  $k_{\text{B}}$ is the Boltzmann constant. To fully explore the parameter space, we carry out optimizations initialized with various simulated annealing temperatures, and potential parameters. 

\begin{table}[b]
  \label{tab:OptPar}
\setlength{\tabcolsep}{4.5pt}
  \begin{tabular*}{0.48\textwidth}{@{\extracolsep{\fill}}*{9}{c}}
    \hline
      & A & n & \(\lambda_{1}\) & \(k_{1}\) & \(\delta_{1}\) & \(\lambda_{2}\) & \(k_{2}\) & \(\delta_{2}\) \\
    \hline
    DIA & 0.34 & 3.40 & 0.73 & 54.13 & 2.77 & 0.61 & 3.72 & 1.08\\
    \hline
    SC & 0.40 & 5.32 & 0.25 & 58.11 & 2.64 &  0.53 & 4.35 & 1.05\\
    \hline
  \end{tabular*}
  \caption{\ Optimal parameters of the potential in equation ~\ref{eq:fomin} for diamond (DIA) and simple cubic (SC) target lattices.}
\end{table}

\begin{figure}[t]
\centering
  \includegraphics[scale=0.54]{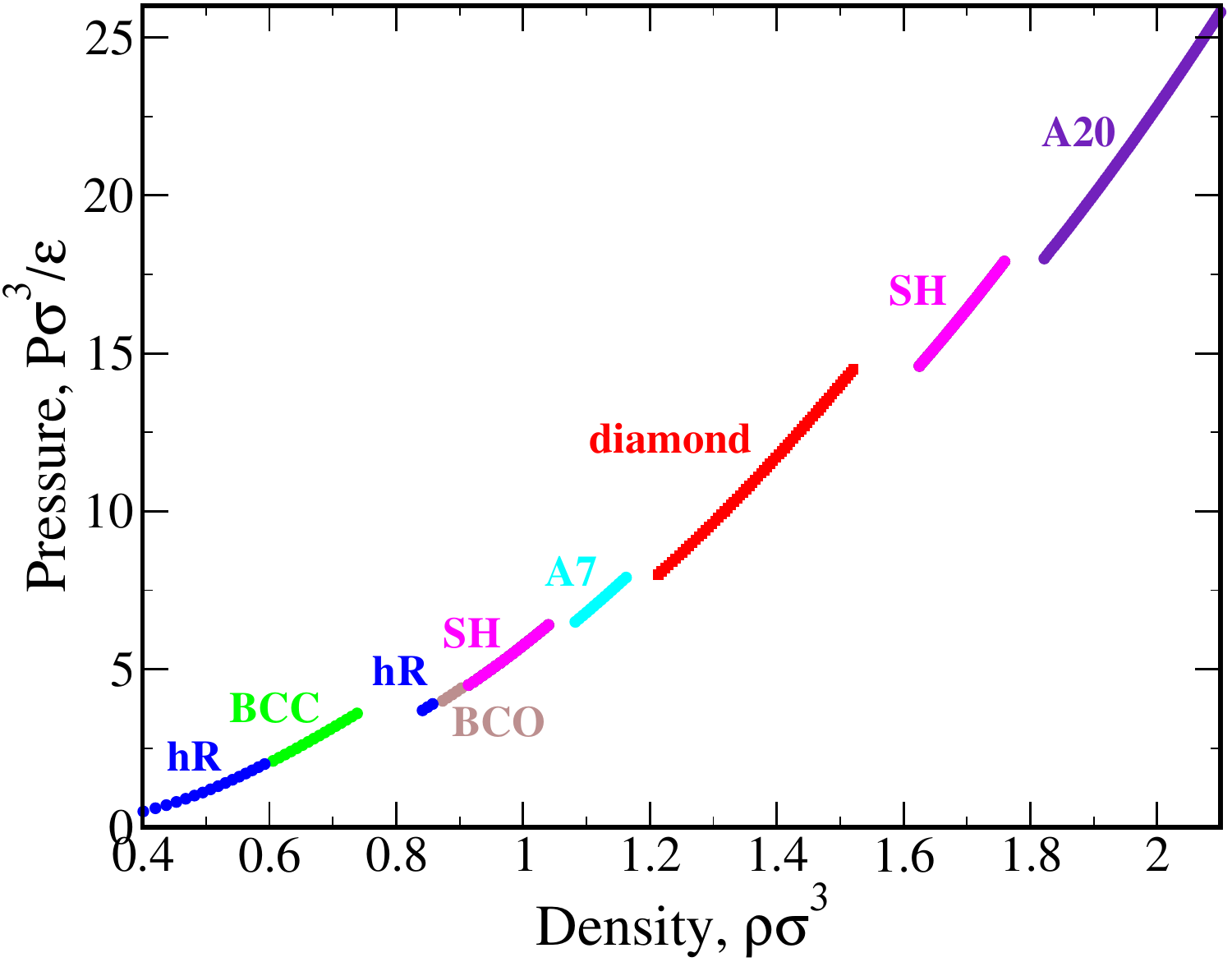}
  \includegraphics[scale=0.54]{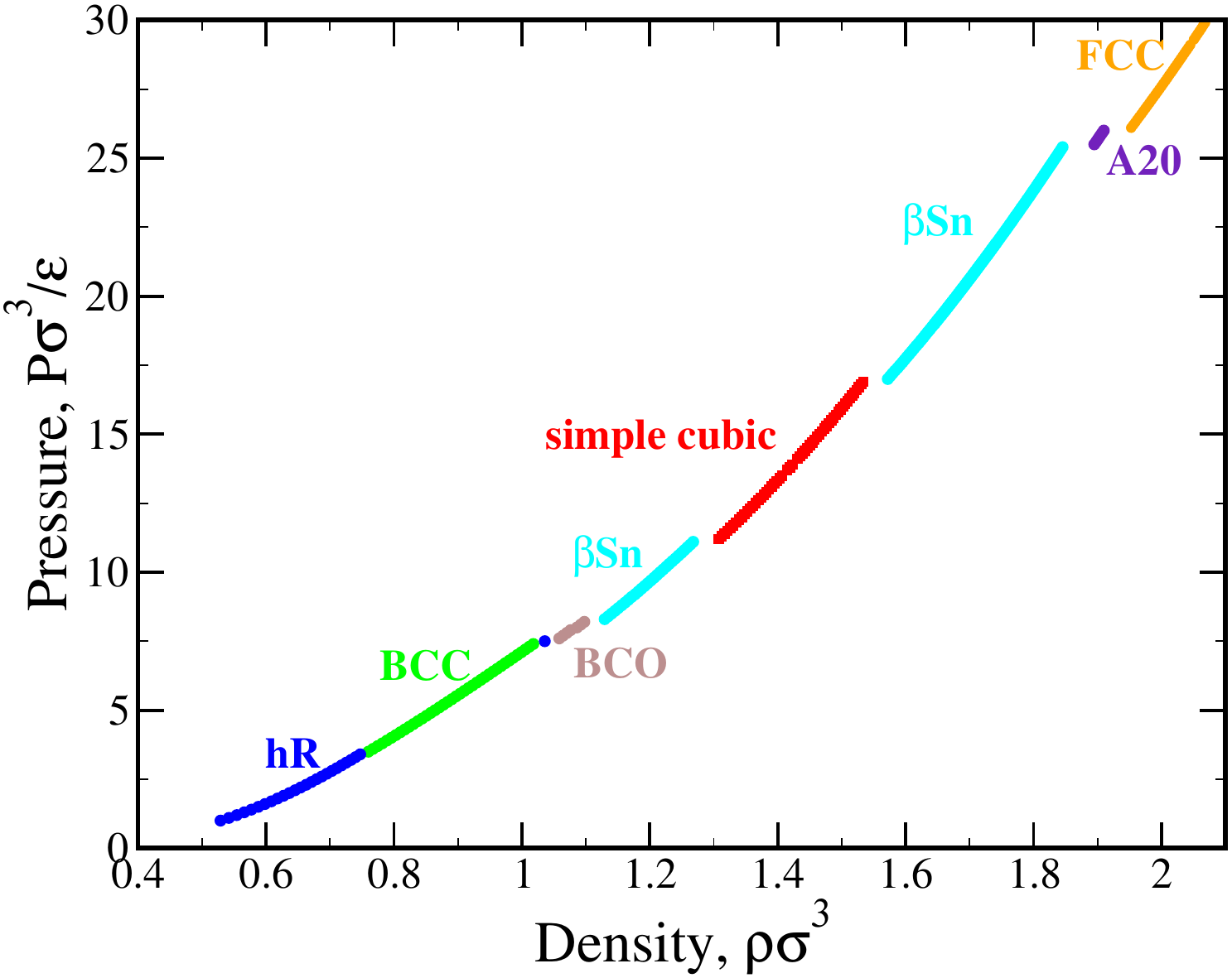}    
  \caption{Ground-state phase diagrams of the optimized potentials (given in Table 1) for - \emph{(top)} a diamond (DIA) target lattice, and \\ \emph{(bottom)} a simple cubic (SC) target lattice. Lattice parameters are reported in the supplementary material$^\dag$.}
\label{fgr:Phasediamond}
\end{figure}

\begin{figure*}[t]
\centering
  \includegraphics[width=6in]{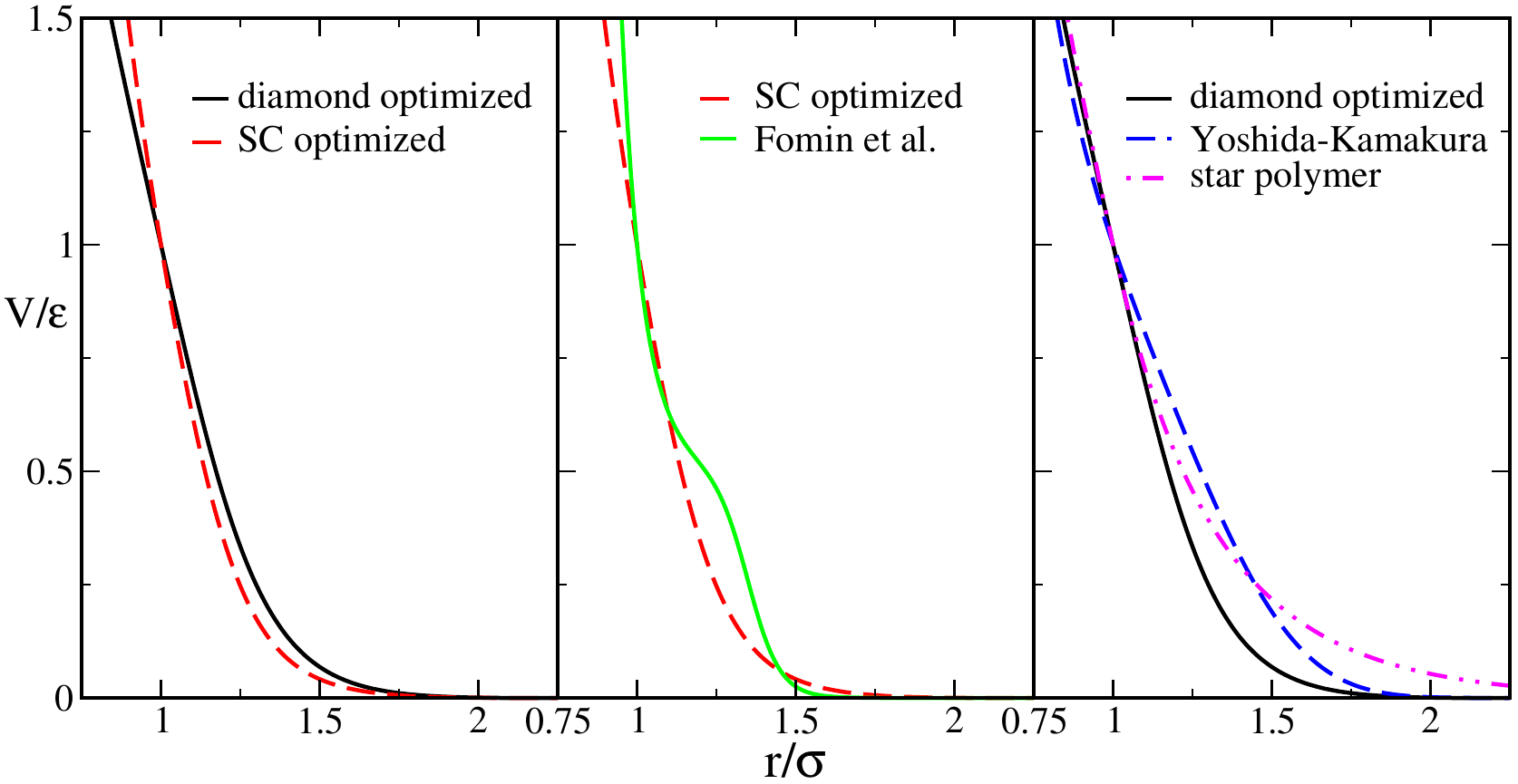}
  \caption{\emph{(left)} Optimized potentials (from Table 1 and equation \ref{eq:fomin}) for diamond and simple cubic (SC) target ground states from this work; \emph{(center)} optimized potential for the SC lattice from this work and a SC-forming potential developed by Fomin et al.\cite{fominsc};\emph{(right)} optimized potential for the diamond lattice from this work compared with an isotropic star polymer interaction model\cite{LikosStarPot} and the Yoshida-Kamakura potential\cite{YKorig1,YKorig2}, which also exhibit diamond ground states.}
\label{fgr:Potentials}
\end{figure*}
 
To ensure success of this optimization strategy, the competitive pool should ideally consist of all lattices which have chemical potentials that are similar to (or less than) that of the target structure for the class of pair potentials and state points under consideration. Motivated by the results of an extensive ground-state study on related models~\cite{zerotempsanti}, we choose lattices for the competitive pool from the following types of periodic structures: face-centered cubic (FCC), body-centered cubic (BCC), diamond (DIA), simple cubic (SC), wurtzite (WUR), hexagonal (SH), body centred orthorhombic (BCO), rhombohedral (hR), A7, A20 and \(\beta\)Sn. Based on extensive preliminary calculations that we carried out for this study--which involved optimizing potential parameters using simulated annealing and computing ground state phase diagrams--we select the following specific lattices in the competitive pools (adopting previously introduced parameter nomenclature\cite{zerotempsanti}) for use when the target lattice is diamond [FCC, WUR, SH \((c/a=1.5)\), \(\beta\)Sn \((c/a=1.39)\), \(\beta\)Sn \((c/a=1.25)\), A7 \((b/a=3.79, u=0.1385)\), A20 \((b/a=1.728, c/a=0.626,y=0.167)\)] and when the target lattice is simple cubic [FCC, BCC, DIA, SH \((c/a=1)\), SH \((c/a=1.08)\), SH \((c/a=1.172)\), A20 \((b/a=1.72, c/a= 0.66, y= 0.67)\), \(\beta\)Sn \((c/a=0.873)\), \(\beta\)Sn \((c/a=0.78)\), \(\beta\)Sn \((c/a=1.75)\)]. Other lattices with different parameters may, of course, turn out to be more stable than these or the target structures under a given set of conditions. To test this possibility, we must carry out a \emph{more extensive ``forward" calculation} of the ground state phase diagram with our final optimized potentials. 

There are a number of sophisticated search routines (e.g. genetic algorithms\cite{GALikos1,MPLikos,bianchiJCP12} and metadynamics~\cite{metadyn}) that have been developed to find the most stable subset of lattices to consider in the forward calculation for a given potential. In this work, we construct the zero-temperature phase diagram for the optimized potential by searching for the most stable lattices from among the periodic structures mentioned above.  For the structures which are defined by lattice parameters (WUR, SH, BCO, hR, A7, A20 and \(\beta\)Sn), we use simulated annealing to obtain the optimal values of these parameters (i.e., those that minimize chemical potential) as a function of pressure. We also verify the mechanical stability of the optimized lattices on the phase diagram by analyzing their phonon spectra. The phase diagram is then constructed from among these energetically and mechanically stable structures with optimized lattice parameters.
%We use simulated annealing to obtain the optimal values of the lattice paramters for 
%Using simulated annealing, for a given interaction potential, we optimize the lattice parameters for these structures as a function of pressure by minimizing the chemical potential of the lattice. 

Parameters of the pair potentials optimized for diamond and simple cubic target structures, respectively, in our simulations are reported in Table 1. The corresponding ground state phase diagrams are shown in Figure \ref{fgr:Phasediamond}. More information on the lattices is provided in the electronic supplementary material$^\dag$. The first point to note is that both optimization strategies are successful in producing their target ground states over a wide density range. The stable density \((\rho\sigma^{\text{3}})\) range for the diamond phase is \(0.308 \; [1.213,1.521]\) and for simple cubic phase is \(0.226 \; [1.308,1.534]\) on the phase diagrams for their respective optimized potentials. These density ranges are considerably larger than those exhibited by the few other published models with isotropic potentials that can display these phases.

 \begin{figure}[t]
  \centering
  \includegraphics[scale=0.35]{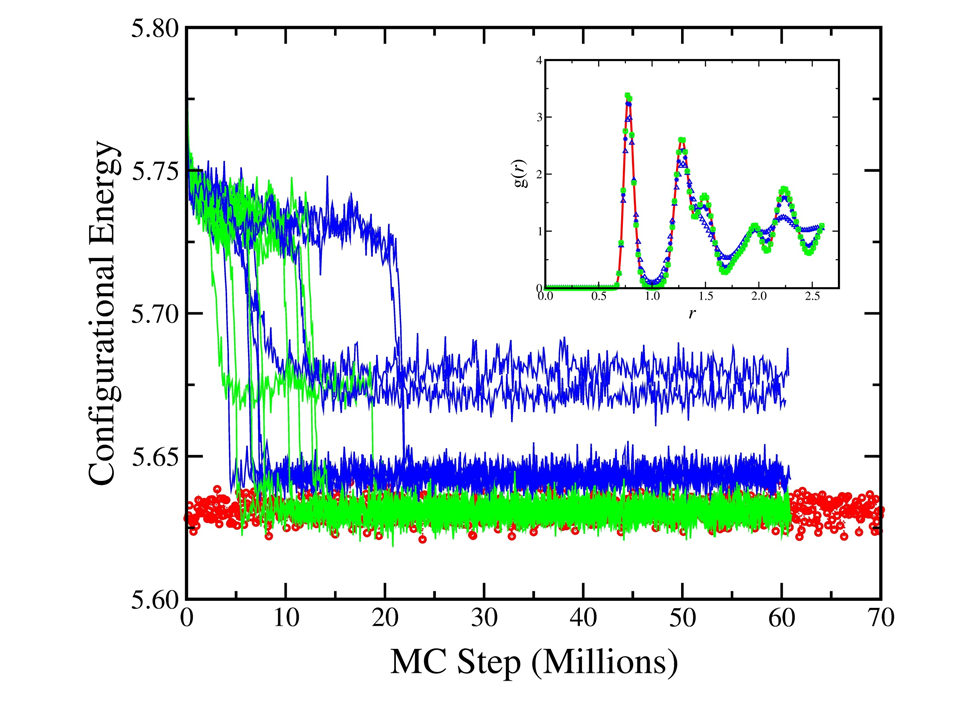}
  \caption{Evolution of the configurational energy during a canonical MC simulation at \(T\)=0.040 and \(\rho\)=1.35. The red circles show the evolution of a system initiated as a perfect diamond lattice.  The blue and green curves (represent a total of 16 configurations initiated with a high-temperature fluid configuration) correspond to configurations that crystallize with and without defects, respectively.  The inset provides the pair correlation functions for some of the final structures. The solid red curve represents the system initialized from a perfect diamond lattice, whereas the blue and green symbols correspond to systems initialized from the high-temperature fluid. Blue and green symbols correspond to configurations that crystallize with and without defects, respectively.}
\label{fgr:MC}
\end{figure}

Two relevant comparisons that can be made for the stability range of the diamond structure are to the coarse-grained center-of-mass star polymer interaction model developed by Watzlawek et al.\cite{LikosStarPot}  and to another model introduced by Yoshida and Kamakura~\cite{YKorig1,YKorig2}. Although neither strictly satisfy all of the ``simplicity" constraints of our optimized model potentials, they are simple no less and have been shown to exhibit stable diamond structures on their phase diagrams. Adopting the same non-dimensional representation of the present study, i.e., \(V(1)=\epsilon\), the star polymer potential~\cite{MPLikos} (molecules with $f=20$ arms  and \(\epsilon=k_B T\)) has a stable diamond density range \((\rho\sigma^{\text{3}})\) of \(0.169\), while the Yoshida-Kamakura~\cite{zerotempsanti} potential has a diamond phase density range of \(0.175\). Both density ranges are roughly half of that exhibited by the potential optimized for the diamond structure in the present study. For the simple cubic structure, there are even fewer relevant comparisons. The one model\cite{fominsc} that we are aware of exhibits a stable simple cubic ground state in a narrow density range of \(0.03\) ($\approx13$\% of the range displayed by the optimized potential presented in this work). For  comparison, we plot the optimized potentials from our study along with the other potential models discussed above in Figure ~\ref{fgr:Potentials}. 

Although the primary focus of the present work is designing target ground states stable over wide density ranges, we have also completed some Monte Carlo (MC) simulations to probe the thermal stability of the diamond-forming system introduced here.  We first completed a series of canonical MC simulations with \(N\) = 250 at \(\rho\) = 1.35 to examine the melting and freezing behavior.  To estimate the melting point, we allowed a diamond lattice to relax at several temperatures separated by \(\Delta T\) = 0.005 (in units of $\epsilon/k_{\text B}$), and found that the diamond lattice melted at temperatures of \(T = 0.075\) and above.  To better understand the assembly process, we allowed a liquid, initially equilibrated at \(T = 1.0\), to relax at several temperatures, and found that the system assembles into a diamond crystal at temperatures of \(T = 0.045\) and below.  Figure ~\ref{fgr:MC} provides data related to this assembly process. Specifically, we show the configurational energy as a function of MC step for a system equilibrated at \(T = 0.040\).  In this case, we find that each of the 16 configurations examined crystallize during the simulation, with 8 of the configurations forming a defect-free diamond lattice and 8 of the configurations assembling into defective diamond crystals.  The nature of the underlying lattice was verified by examining the pair correlation functions. Similar results were obtained with a system consisting of \(N = 1024\) particles.  Collectively, these data suggest that the diamond system exhibits a first-order melting transition at \(\rho\) = 1.35.

\begin{figure}[!b]
  \centering
  \includegraphics[scale=0.35]{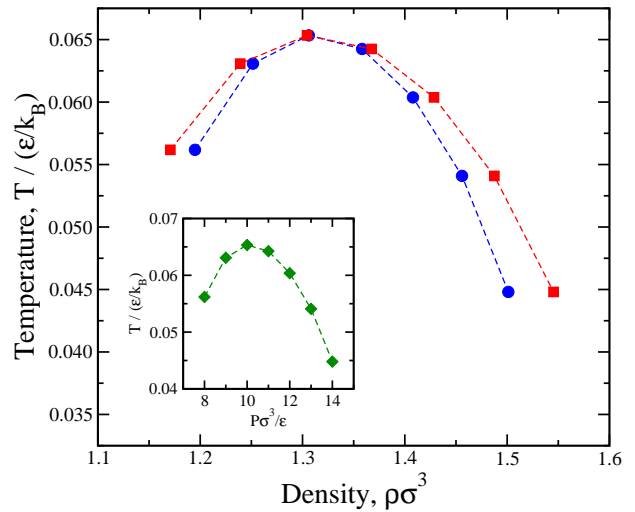}
  \caption{Phase diagram for the diamond-forming potential.  The main panel provides the temperature-density plane.  Circles and squares provide the saturated densities of the diamond and fluid phases, respectively.  The inset provides the phase diagram within the temperature-pressure plane.  Dashed lines are simply guides to the eye. The statistical uncertainty of the simulation data is smaller than the symbol size.}
\label{fgr:FiniteTempPD}
\end{figure}

We are now employing free energy MC methods to construct phase diagrams for the systems introduced here.  Figure ~\ref{fgr:FiniteTempPD} provides initial data related to the diamond-fluid saturation curve.  These points were located by finding the temperature at which the Gibbs free energy of the fluid matched that of the diamond crystal along a given isobar.  The temperature dependence of the fluid's Gibbs free energy was computed via a combination of grand canonical transition matrix MC~\cite{JeffGCMC} and isothermal-isobaric temperature expanded ensemble MC~\cite{JCP1992} simulations.  The temperature dependence of the crystal's Gibbs free energy was computed via a combination of Frenkel-Ladd MC~\cite{FLMC} and isothermal-isobaric temperature expanded ensemble MC~\cite{JCP1992} simulations.  Our results point to a concave-shaped diamond-fluid saturation curve within the temperature-pressure plane, with the temperature maximum located at approximately \(T = 0.065\), where \(P = 10\) and \(\rho\) = 1.31.  The heating and cooling simulations outlined above were completed at a density slightly beyond this maximum point.  The Yoshida-Kamakura system exhibits a similarly-shaped diamond-fluid saturation curve with a lower maximum melting temperature~\cite{finitetempYK} of \(T = 0.047\).  These results suggest that the thermal stability of the current model exceeds that of the Yoshida-Kamakura model.

To summarize, our investigation shows that it is possible to use techniques of inverse statistical mechanical optimization to obtain simple pairwise interaction forms with targeted low-coordinated three-dimensional structures stable over a wide density range. We will be presenting a detailed study of the Monte Carlo free-energy simulation methods to compute the thermal stability (i.e., the temperature-dependent phase diagrams) of the systems introduced here in a future publication. Additionally, we plan to study whether a systematic coarse-graining strategy (e.g., relative entropy maximization~\cite{shellJCP2008}) could preserve low-coordinated ground states when mapping from anisotropic ``patchy" interactions to simpler, isotropic effective potentials.

As a final note, after finishing this manuscript, we became aware of a very recent preprint by Marcotte et al.\cite{arXiv} which also reports a convex-repulsive pair potential that exhibits a diamond ground state. They use a different inverse statistical mechanical optimization method than that reported in this study with and obtain a considerably narrower range of thermodynamic stability. However, the resulting interaction potential and its derivatives, although of different functional forms, are strikingly similar to those we report here, providing further confirmation of the robustness of the qualitative result.

T.M.T. acknowledges support of the Welch Foundation (F-1696) and the National Science Foundation (CBET-1065357). J.R.E. acknowledges support of the National Science Foundation (CHE-1012356).  We also acknowledge the Texas Advanced Computing Center (TACC) at The University of Texas at Austin and the Center for Computational Research at the University at Buffalo for providing HPC resources that have contributed to the research results reported within this paper.

\balance

\footnotesize{
\bibliography{Ground_state} %your .bib file
\bibliographystyle{rsc} 
}

\end{document}